\documentclass[aps,10pt,onecolumn,amssymb,amsmath,nofootinbib,floatfix,
notitlepage]{revtex4-1}

\usepackage{graphicx}
\usepackage{subfig}

\newcommand{\del}{\ensuremath{\delta}}

\newcommand{\Lam}{\ensuremath{\Lambda}}
\newcommand{\sig}{\ensuremath{\sigma}}

\newcommand{\Msun}{\ensuremath{M_{\odot}}}

\newcommand{\Mz}{\ensuremath{(M,z)}}
\newcommand{\grmgrz}{\ensuremath{\tilde{R}_{>M>z}}}
\newcommand{\nbgrmgrzi}{\ensuremath{{R}^M_{i>M>z}}}
\newcommand{\nbgrmgrz}{\ensuremath{{R}_{>M>z}}}
\newcommand{\nblessf}{\ensuremath{{R}_{<f}}}
\newcommand{\nblessfi}{\ensuremath{{R}^M_{i<f}}}
\newcommand{\Cnblessf}{\ensuremath{{R}^C_{i<f}}}
\newcommand{\Hgrmgrz}{\ensuremath{\tilde{R}^H_{i>M>z}}}
\newcommand{\fsky}{\ensuremath{f_\mathrm{sky}}}

\newcommand{\fig}[1]{Fig. \ref{#1}}

\newcommand{\ph}[1]{\phantom{#1}}
\newcommand{\tab}[1]{Table \ref{#1}}

\newcommand{\be}{\begin{equation}}
\newcommand{\ee}{\end{equation}}
\newcommand{\la}[1]{\label{#1}}

\newcommand{\fnl}{\ensuremath{f_{\rm NL}}}
\newcommand{\gnl}{\ensuremath{g_{\rm NL}}}

\begin{document}
\title{Quantifying the rareness of extreme galaxy clusters}
\author{Shaun Hotchkiss}
\email{shaun.hotchkiss@helsinki.fi}
\affiliation{Department of Physics, University of Helsinki and
  Helsinki Institute of Physics, P.O. Box 64, FIN-00014 University of
  Helsinki, Finland}
\begin{abstract}
I show that the most common method of quantifying the likelihood that an extreme
galaxy
cluster could exist is biased and can result in false claims of tension with
\Lam CDM. This common method uses the probability that at least one cluster
could exist above the mass and redshift of an observed cluster. I demonstrate
the existence of the bias using sample cluster populations, describe its origin
and explain how to remove it. I then suggest potentially more suitable and
unbiased measures of the rareness of individual
clusters. Each different measure will be
most sensitive to different possible types of new physics. I show how to
generalise these measures to quantify the total `rareness' of a set of clusters.
It is seen that, when mass uncertainties are
marginalised over, there is no tension between the standard \Lam CDM
cosmological model and the existence of any
observed set of clusters. As a case study, I apply these rareness
measures to sample cluster populations generated using primordial density
perturbations
with a non-Gaussian spectrum.
\end{abstract}
\maketitle
\section{Introduction}

Observations of galaxy clusters with sufficiently large masses, at sufficiently
large redshifts, can provide strong constraints on possible deviations from the
standard
\Lam CDM cosmological model with Gaussian primordial
perturbations~\cite{Allen:2011zs}. Even the observation of a
\emph{single}, sufficiently `big' and `early' cluster could rule out the
standard model with a high level of confidence~\cite{Mortonson:2010mj}.
Furthermore, a number of clusters have recently been measured to have masses and
redshifts
\cite{Mullis:2005hp,Broadhurst:2008re,Mantz:2009fx,Richard:2009sv} that do, even
individually, seem to create tension with the combination of \Lam CDM and
Gaussian initial perturbations~\cite{Jee:2009nr,Holz:2010ck,Foley:2011jx}. When
attempts have been made to quantitatively measure the degree to which these
clusters are collectively ``too big, too early'' large discrepancies between
\Lam
CDM and the clusters have been
claimed~\cite{Hoyle:2010ce,Enqvist:2010bg}. More of these high mass and high
redshift clusters have subsequently been detected by the South Pole Telescope
(SPT) \cite{Williamson:2011jz} and many more will be detected in the near future
by SPT, the Planck satellite, the Dark Energy Survey and the XMM Cluster Survey
(XCS)~\cite{LloydDavies:2010ue}.

The formation of extreme galaxy clusters depends on both the size of the
primordial density perturbations and how long these perturbations have had to
grow. In the simplest theoretical models, clusters form when $\delta_R$, which
is the fractional density contrast,
$\delta$, averaged over an approximately spherical region of space with radius
$R$, crosses a particular threshold. Due to the assumption of statistical
homogenity, at a given time, every point in space has the same
probability distibution for $\delta$. Therefore, when $\delta$ is averaged over
larger volumes, the fluctuations in $\del_R$ become smaller. This means that
regions with large radii are less likely to collapse than
regions with small radii. Unfortunately, after collapse it is not possible to
directly measure the radius of the region that collapsed to form a cluster.
Nevertheless, according to theory, every cluster forms when $\delta_R$ crosses
one definite threshold. This necessitates that a cluster with a larger mass came
from a larger region of space. Therefore, it is expected that fewer clusters
will form at larger masses. Fewer clusters are also expected to form at higher
redshifts simply because the initial density contrast has had less time to grow.
Therefore, the masses and redshifts of observed clusters are good parameters for
quantifying the rareness of extreme galaxy clusters. As a result, if it could be
shown that the most extreme clusters did form ``too big, too early'', or even
``too little, too late'', it would be possible to conclude that either
the primordial perturbations do not have a Gaussian spectrum with scale
dependence given by a simple power or that the calibration between time and
redshift is wrong and therefore clusters did not form as early as is inferred
from
\Lam CDM.

It would seem that a very natural method to quantify the rareness of an extreme
galaxy cluster is to ask ``what is the probability of observing at least one
cluster at this mass and redshift or above ($>M>z$)?'' Once uncertainties in
mass
measurements, selection functions and cosmological parameters are properly
marginalised over this question can then be
converted into a measure of the probability that at least one cluster this rare
could be \emph{observed}. It is this resulting measure that has been used to
claim that the existence of certain clusters provides tension
with \Lam CDM and a Gaussian, power law, spectrum of density perturbations. This
measure is biased. When rareness is measured without bias it is seen that, in
\Lam CDM, there is a greater than 50\% probability that the rarest currently
observed cluster could exist somewhere in the fraction of sky observed.
Similarly, when the combined rareness of a set of clusters is calculated
correctly, it is seen that there is a greater than 20\% probability that even
the entire set of rarest observed clusters could exist. Nevertheless it is still
true that quantifying the rareness of extreme galaxy clusters might provide
useful information about possible new physics as will be explored in Section
\ref{sec:nogauss}.

The core of this paper is structured as follows. In Section \ref{sec:>M>z}, I
will discuss the $>M>z$ measure of rareness. I will
discuss why to expect that this measure will give biased results.
Following this, in Section \ref{sec:>M>zsim}, I will show explicitly that it
does lead to biased results when applied to sample cluster populations that were
generated assuming a \Lam CDM universe. In Section \ref{sec:nbsingclust}, I will
show how to alter this measure to remove the bias. I will also
present other well motivated measures to quantify the rareness of individual
clusters. Different measures will be more useful depending on what specific
deviations from \Lam CDM are being tested. In Section \ref{sec:manyclust}, I
will
introduce measures that combine the individual rarenesses of a set of clusters
to quantify the total rareness of the set as a whole. In both Section
\ref{sec:nbsingclust} and Section \ref{sec:manyclust}, I will apply the
introduced measures of rareness to sets of observed galaxy clusters. Finally, in
Section \ref{sec:nogauss}, I will apply these rareness measures to sample
cluster populations generated from non-Gaussian primordial perturbations.

\subsection{The expected number of clusters in a region of the $(M,z)$
plane}\la{sec:exp}

Before moving on to the core of the paper I will briefly outline the methodology
used to calculate the expected number of clusters in a given region of mass and
redshift over a given fraction of the sky, $f_\mathrm{sky}$. For each measure of
rareness I will be calculating the probabilities of clusters existing assuming a
Poissonian distribution; therefore only the expected number is needed. This
quantity is
given by the following integral
\begin{equation}\label{eq:expected}
\mathrm{N}=f_\mathrm{sky}\int dz \frac{dV}{dz} \int dM  \frac{dn(M,z)}{dM}
 \end{equation}
where the $z$ and $M$ integrals are over the region of the $(M,z)$ plane being
considered, $dn/dM$ is the number density of galaxy clusters (the halo mass
function) and $dV/dz$ is the volume element,
\begin{equation}
\frac{dV}{dz}=\frac{4\pi}{H(z)}\left[\int^z_0 \frac{dz'}{H(z')}\right]^2.
 \end{equation}
It is customary to reparameterise $dn/dM$ as
\begin{equation}
 \frac{dn}{dM}=\frac{\rho_{m}}{M^2} \frac{d\ln\sigma(M,z)^{-1}}{d\ln M}
f(\sigma,z)
\end{equation}
where $\rho_{m}$ is the \emph{present} density of matter in the universe and
\sig~is the variance of the density contrast smoothed over the comoving scale,
$R$, that corresponds to the mass, $M$. The
function $f$ (often also called the mass function, a convention I will follow in
this work) must then either be calculated theoretically or by matching to
simulations. The form of $f$ depends on the shape of the primordial
spectrum. For most of this work I will be assuming that the primordial spectrum
is Gaussian and will use the Tinker et al. mass function~\cite{Tinker:2008ff},
\begin{equation}
 f(\sigma,z)=A\left[
\left(\frac{\sigma}{b}\right)^{-a}+1\right]e^{-\frac{c}{\sigma^2}}
\end{equation}
with $A=0.186(1+z)^{-0.14}$, $a=1.47(1+z)^{-0.06}$, $b=2.57(1+z)^{-0.011}$ and
$c=1.19$. I will calculate $\sigma(M,0)$ using the fit to the transfer function
presented in \cite{Bardeen:1985tr} with the modified shape parameter introduced
in \cite{Sugiyama:1994ed}. The redshift dependence of $\sigma$ is then
obtained by multiplying $\sigma(M,0)$ by the linear growth function normalised
to be proportional to $1/(1+z)$ during the era of matter domination. Unless
otherwise stated I will also use the WMAP 7 Cosmology maximum likelihood (ML)
parameters~\cite{Komatsu:2010fb}. Using a different set of cosmological
parameters (specifically $\sigma_8$) would change the exact numbers quoted in
parts of this paper but would not alter any of the conclusions drawn from these
numbers.

\subsection{Notation used to classify different rareness measures}

In this work I will be introducing a number of different measures to quantify
cluster rareness. In each case, the quoted measure is to be interpreted as the
probability that at least one cluster (or set of clusters) could exist that is
at
least as rare as the measured cluster (or set). For reference, I will quickly
summarise the notation convention I have used. For every measure I use $R$ as
the base for an unbiased measure and $\tilde{R}$ for a biased measure. For
individual cluster rareness, each measure corresponds to a prescription for
defining unique contours in the $(M,z)$ plane. I have chosen to depict the
contour used for each measure with subscript text. For example, for the biased
version of the $>M>z$ measure of rareness I write $\tilde{R}_{>M>z}$. I also
introduce a number of ways to combine the individual rarenesses of a set of
clusters. I have chosen to distinguish each method with superscript text. For
example, for the method used by Hoyle et al. in Ref.\cite{Hoyle:2010ce} I
write $\tilde{R}^H_i$. To depict the number of clusters used in a combined
measure I put this number in subscript immediately before the contour
definition. In principle, for a set of clusters, it is possible to combine any
contour definition with any method for combining the individual cluster
rarenesses.

\section{\grmgrz~as a measure of rareness}\label{sec:>M>z}
\subsubsection{Why there is a bias}
To quantify the rareness of a cluster using the $>M>z$ measure it is necessary
to first calculate the number of clusters expected above the mass and redshift
of the observed cluster. This expected number is then
converted into the probability that at least one cluster at least as rare as the
observed cluster could exist, $\grmgrz$, by assuming Poissonian statistics. If
this probability is small, the cluster is deemed to be rare and tension can be
claimed with \Lam CDM.

The bias in this measure can be seen in the following way. Using the definition
given it is possible to define a value of \grmgrz~at every point in the $(M,z)$
plane. It is then possible to construct contours of equal \grmgrz~in the plane.
The number of clusters expected above each contour of constant \grmgrz~is
\emph{necessarily larger} than the expected number of clusters used to calculate
\grmgrz. This is because, for every point in the plane, \grmgrz~ is calculated
using \emph{only} the expected number of clusters at greater masses \emph{and}
greater redshifts. Nevertheless, there will be points at lower masses and
greater redshifts (and vice versa) that lie on the same \grmgrz~contour. These
points would not be included in the integration region used to define \grmgrz~at
any other point on the contour, but are defined to be equally rare. Therefore,
for a given $\grmgrz=\tilde{R^*}$, the probability that some cluster exists with
$\grmgrz\leq\tilde{R^*}$ is \emph{necessarily greater} than $\tilde{R^*}$ itself.
This is the bias. This argument is illustrated in
Figure \ref{fig:grmgrz}.
\begin{figure}
\centering
\includegraphics[height=0.25\textheight]{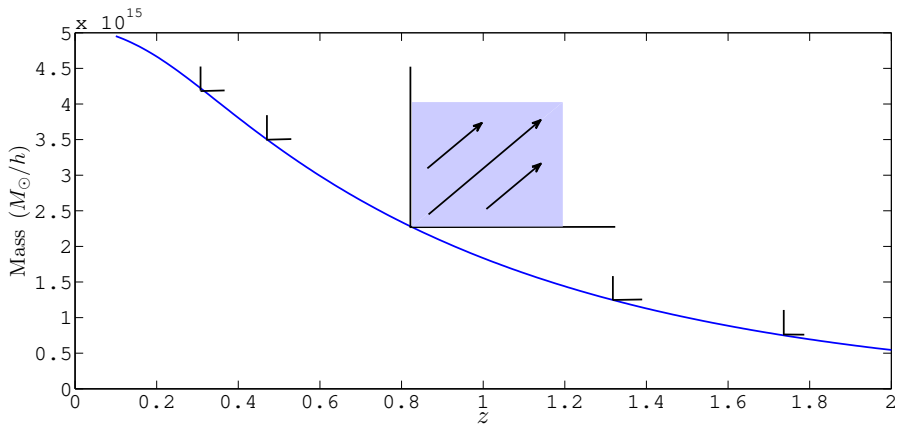}
\caption{\small An outline of the reason for the bias in $\tilde{R}_{>M>z}$.
Every point on the curve has the same number of clusters expected at greater
masses and redshifts (i.e. the same $\grmgrz=\tilde{R^*}$). The probability that a
cluster will exist with $\grmgrz\leq\tilde{R^*}$ is equal to the probability that
a cluster will exist above this entire curve. This probability is necessarily
greater than $\tilde{R^*}$.} 
\la{fig:grmgrz}
\end{figure}
\subsubsection{The magnitude of the bias}
\begin{figure}
\centering
\includegraphics[height=0.25\textheight]{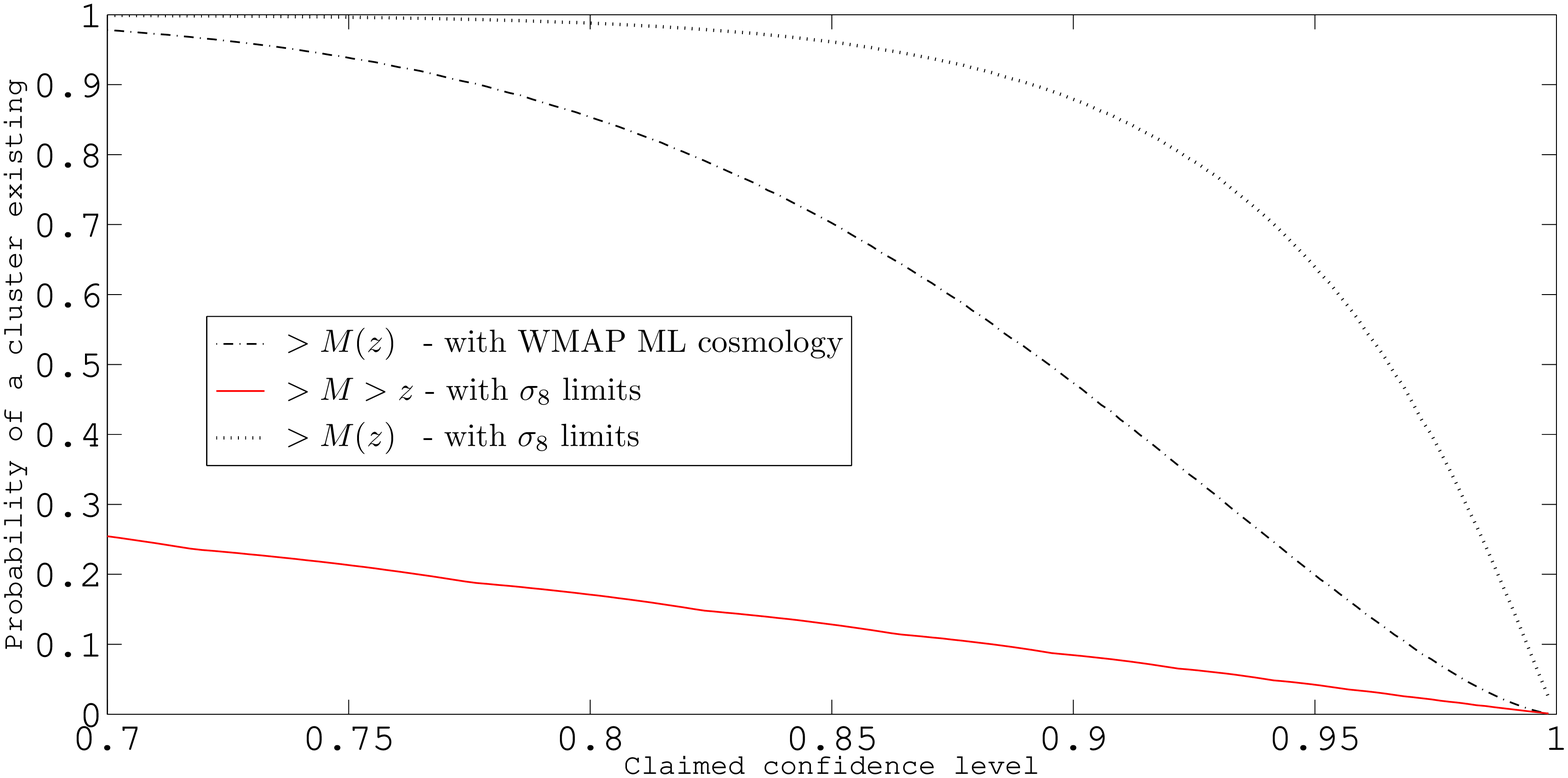}
\caption{\small The solid red line is \grmgrz, the probability of at least one
cluster existing at a greater
mass and redshift than each point on a contour, $M(z)$. The dashed and dotted
black lines are the probability of at least one
cluster existing above the whole $M(z)$ contour. For each data point in the
figure, $M(z)$ is defined so that the value of $1-\grmgrz$ on $M(z)$ is equal to
the value on the x-axis. To produce this figure I have used the fitting formula
for $M(z)$ provided in Ref.\cite{Mortonson:2010mj}. See the main text for a
discussion of which cosmological parameters were used for each line.} 
\label{fig:mortplot}
\end{figure}

In Refs.\cite{Holz:2010ck,Mortonson:2010mj} the idea of a contour of constant
\grmgrz~is used explicitly. Ref.\cite{Mortonson:2010mj} provides fitting
formulae for the
contours as a function of the exclusion limit required to rule out \Lam CDM.
But, for a given exclusion limit of $\alpha$, if
\emph{any} cluster detected above the corresponding contour is to be interpreted
as ruling out
\Lam CDM with $100\alpha\%$ confidence, then the probability of observing such a
cluster cannot
be greater than $1-\alpha$. In Figure \ref{fig:mortplot} I have tested
this. Ref.\cite{Mortonson:2010mj} makes a distinction between $p$, the
proportion of WMAP 7 parameter space ruled out, and $s$, the probability
to have zero clusters $>M>z$ in a random sample of the same sky fraction.
I have always taken $s=p$ and have used this as the x-axis of Figure
\ref{fig:mortplot}. The y-axis gives the probability of at least one cluster
existing in a specified region of the $(M,z)$ plane, using a specified set of
cosmological parameters. The three different lines correspond to particular
regions and particular cosmological parameters. For each line I have taken
$\fsky=1$.

The solid red line uses the region of $(M,z)$ defined as $>M>z$ of
any arbitrary point on the corresponding contour (by definition all points on
the same contour return the same result). Ref.\cite{Mortonson:2010mj} states
that, for a contour
with a given $p$, in $100p\%$ of allowed cosmological parameter space the
expected number of clusters is less than the expected number that would give the
required statistical exclusion limit, $s$. The expected number of clusters
depends sensitively and monotonically on $\sigma_8$ but very weakly on all other
cosmological parameters. Therefore, for the solid red line I calculate the
expected number of clusters $>M>z$ using
$\sigma_8(p)$ defined by
\begin{equation}\label{eq:mortsig8}
 P(\sigma_8<\sigma_8(p))=p.
\end{equation}
Following Ref.\cite{Mortonson:2010mj}, the distribution for $\sigma_8$ was taken
from combined CMB, SN, BAO and $H_0$
constraints. The red line matches well to expectations indicating that I am
correctly interpreting Ref\cite{Mortonson:2010mj}. The dashed and dotted black
lines correspond to the probability of observing at least one cluster above the
whole $M(z)$ contour. The dashed line comes from using the WMAP maximum
likelihood cosmological values and
the dotted line comes from using $\sigma_8$ defined by eq.(\ref{eq:mortsig8}).
From the dashed line it is seen that in a \emph{true} \Lam CDM universe there is
a 40\% chance of a cluster existing with a value of \grmgrz~that would claim to
exclude \Lam CDM at
90\% confidence. The dotted line makes it possible to estimate the true
conservative significance that should
be claimed by the observation of a cluster on or above a particular $M(z)$
contour. When the dotted line gives a probability of 0.1, the x-axis value is
$\gtrsim 0.99$. Therefore, if a cluster were detected that should result in a
claim of exclusion at only $\sim90\%$ significance, the \grmgrz~statistic would
erroneously claim this as $\gtrsim99\%$ significance.
\subsection{The bias in practice}\label{sec:>M>zsim}
In this section I will examine this bias using sample cluster populations,
generated assuming a \Lam CDM universe. In observations the $>M>z$ contours were
not set until the clusters were observed. Therefore, to simulate the
observations correctly, it is crucial that the contours are set \emph{after} the
sample is generated, at the masses and redshifts of the sample clusters and not
at the masses and redshifts of the clusters observed in our universe.
\subsubsection{Individual clusters}
In this subsection I will calculate \grmgrz~for the most extreme clusters of
sample populations. To generate
the samples I have first broken the $(M,z)$ plane into bins of equal spacing in
$\Delta z$ and $\Delta (\ln M)$. I have then calculated the expected number of
clusters to exist in each bin, assuming \Lam CDM and using the methodology of
Section \ref{sec:exp}.
I then generate a Poisson sample in every bin. The occupied bins determine the
masses and redshifts of a sample  population of clusters. The bin spacing is
ensured to be fine enough such that no bin is
ever occupied by more than one cluster.

For every bin in the \Mz~plane I also calculate \grmgrz. In each sample, I then
order
the occupied bins with respect to \grmgrz. The right hand panel of
\fig{oneclust} is a histogram showing the proportion of times that each
\grmgrz~value was the smallest \grmgrz~value in all of the occupied
bins\footnote{For this figure I took $\fsky=2500$ sq. degrees and imposed
$M>10^{15}\Msun$ (see Section \ref{sec:apploneclust}).}. If \grmgrz~was an
unbiased statistic this would
be uniform\footnote{$R$ is supposed to quantify the probability that \emph{any}
cluster could exist that is at least as rare as a given cluster. Put another
way, for an unbiased measure, if $R^*$ is the rareness of the rarest cluster
then $P(R^*\leq R)=R$. This is the definition of a uniform distribution.};
instead, smaller values of \grmgrz~are
disproportionately favoured. The histogram in the left hand panel is an
identically
produced histogram for one of the unbiased statistics that will be introduced in
Section \ref{sec:unbiased}.
\begin{figure}
\centering
\includegraphics[height=0.25\textheight]{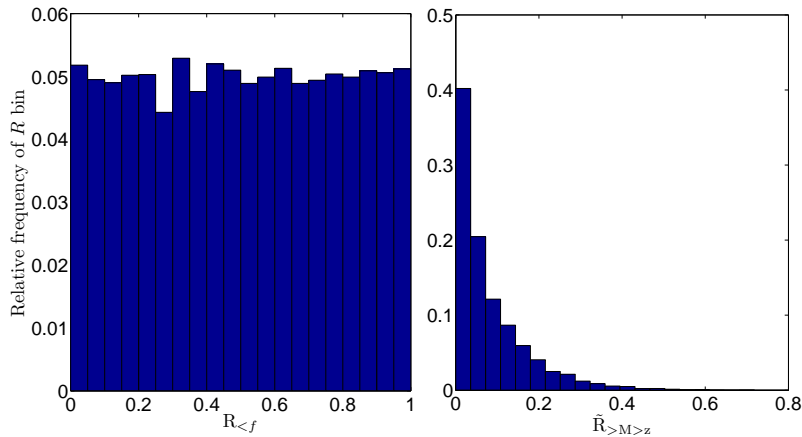}
\caption{\small Histograms showing the distributions generated for an unbiased
rareness measure (left panel) and a biased measure (right panel) when they are
applied to sample cluster populations. The x-axis on the right panel is the probability that a
cluster can exist at greater mass and redshift than the rarest cluster in each
sample.} 
\la{oneclust}
\end{figure}
\subsubsection{Multiple clusters}\label{sec:multclustbias}
In Refs.\cite{Hoyle:2010ce,Enqvist:2010bg} the individual \grmgrz~values for a
set of
clusters were combined to form an estimate, \Hgrmgrz, of the rareness of the set
as a whole.
This was done simply by multiplying together the \grmgrz~values for
each of the individual clusters. These references analysed a particular
table of clusters put together in Ref.\cite{Hoyle:2010ce} (hereafter referred to
as H10). The table consisted, at the time,
of all the clusters that had been detected, with spectroscopic follow up, above
$z=1$. The masses, redshifts and mass errors for the table of clusters are
quoted in Table 1 of both references. Some clusters in this table were detected
by X-ray experiments and some from SZ experiments. The convention used in the
references was to take
$f_\mathrm{sky}=283$ sq. degrees for the X-ray detected clusters and
$f_\mathrm{sky}=178$ for the SZ detected clusters. To compute \Hgrmgrz~for H10
it is necessary to account for mass uncertainties. Both references assumed that
the masses of the clusters in the table were
log-normally distributed. They then sampled from the mass distribution for each
cluster and, for each sample, calculated the corresponding \Hgrmgrz~statistic.
The final value of \Hgrmgrz~for the H10 clusters is the mean. The result is very
small, $\sim 10^{-3}$, and resulted in claims of large tension.
\begin{table}
\begin{tabular}{|c|c|c|}\hline
 \ph{hi0} Cluster set \ph{hi0} & \ph{0000}Mean \Hgrmgrz\ph{0000} &
\ph{000}Median \Hgrmgrz\ph{000} \\ [1ex]\hline \hline
 H10 & $6.1\times10^{-3}$ & $2.6\times10^{-6}$ \\ [0.5ex]\hline
 Sample (rarest) & $3.0\times10^{-4}$ & $1.5\times10^{-6}$ \\ [0.5ex]\hline
 Sample (random) & $0.70$ & $0.84$ \\ [0.5ex]\hline
\end{tabular}
\caption{\small The measure of rareness used in H10 (i.e.
Ref.\cite{Hoyle:2010ce}), applied
to the table of clusters in H10 and to sample \Lam CDM cluster
populations. This measure was used in \cite{Hoyle:2010ce,Enqvist:2010bg} to
claim significant tension with \Lam CDM}   
\label{tab:Hoyle}
\end{table}

I have repeated this calculation in \tab{tab:Hoyle} with two minor changes.
Firstly, I
have removed the lightest cluster from H10. This has no effect
on the result. Secondly, I have used $\fsky=178+283=461$ sq. degrees for all
clusters. This does
affect the quoted numbers but not the conclusions that can be drawn from them.
This choice was made because it makes it simpler to compare to
sample populations\footnote{It is also arguably the more correct choice.
However, treating the SZ and X-ray \fsky~values separately is not clearly
\emph{wrong} because not all of the X-ray detected clusters would have been
detected by an SZ experiment.}, which I have also done. To generate the samples
I made a cut of $z>1$ and $M>2\times
10^{14}\Msun$ to attempt to imitate the nature in which H10 was constructed.
Including clusters in the sample below either of
these thresholds would increase the effects of the bias by increasing the
number of clusters with small \grmgrz. In Table \ref{tab:Hoyle}, the sample
(rarest) entry corresponds to the \Hgrmgrz~value calculated using the 13
occupied bins with the smallest \grmgrz. The sample (random) corresponds to the
\Hgrmgrz~value calculated using 13 random occupied bins. It is clear that the
mere existence of
the H10 clusters does not provide tension with \Lam CDM.
\begin{figure}
\centering
\includegraphics[height=0.25\textheight]{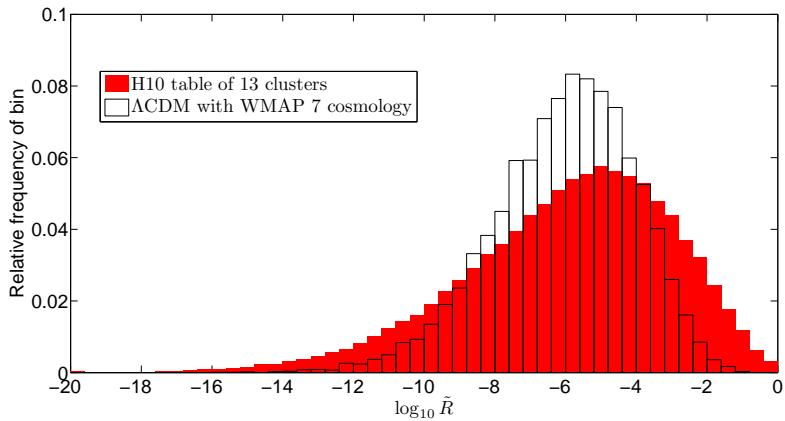}
\caption{\small Histograms of the measure of rareness introduced in H10 applied
to the table of clusters in H10 and to sample \Lam CDM cluster
populations. The distribution for H10 comes from sampling the mass uncertainties
(see text).} 
\la{fig:hoyleplot}
\end{figure}
In \fig{fig:hoyleplot}, I have plotted histograms for H10 and
the sample populations. For the samples, the distribution plotted is the
proportion of times that a sample had the binned \Hgrmgrz~value using the 13
rarest clusters. For H10, the distribution arises through the log-normal
sampling of the mass errors. The distributions are very consistent with each
other. In fact, an actual \Lam CDM universe would be consistent, to $\sim95\%$
confidence, with values of \Hgrmgrz~as small as~$10^{-10}$.

Due to the conservative methods used to construct H10,\footnote{See
Ref.\cite{Hoyle:2010ce} for full details, but an example is the already
mentioned exclusion of all observed clusters without spectroscopic follow up.}
it is unlikely that these clusters \emph{are} the 13 rarest clusters in this 461
sq. degree window. If these clusters were picked at random from the sky they
\emph{would} be anomalously rare. However, this also does not reflect the
reality. Larger mass clusters are more likely to be seen and the rarest clusters
are more interesting and thus more likely to have been spectroscopically followed
up. If the full H10 selection function were properly taken into account, the
true result for the samples would therefore lie somewhere between the two
presented in Table \ref{tab:Hoyle}. This is exactly where the observed value
lies.

In Refs.\cite{Hoyle:2010ce,Enqvist:2010bg} the apparent tension with \Lam CDM
was used to claim detections of minimum values for the non-Gaussianity
parameters \fnl~and \gnl. Clearly, without taking selection functions into
account, these clusters present no evidence for non-Gaussianity and an unbiased
analysis would return results consistent with $\fnl=\gnl=0$.

\section{Unbiased measures of rareness}\label{sec:unbiased}

It is easy to test whether any measure of rareness is unbiased by applying it
to sample cluster populations. If it returns a uniform distribution, it is
unbiased. All of the unbiased measures in this section have been tested in this
way. An example of this is shown in the left panel of Figure \ref{oneclust}.
This was calculated for \nblessf, however any unbiased measure of rareness would
give an identical result.
\subsection{Single cluster measures}\label{sec:nbsingclust}
\subsubsection{\nbgrmgrz}
The bias in \grmgrz~can be removed. This is done simply by finding, for each
observed value of \grmgrz, the true probability of observing this value of
\grmgrz~or less. This amounts to calculating the expected number of clusters
above a contour, $M(z)$, of constant \grmgrz~and then using \emph{this}
expectation value to calculate, \nbgrmgrz, the probability of at least one
cluster at least this rare existing.

In Ref.\cite{Foley:2011jx} a cluster was detected that was claimed, using
\grmgrz, to have
only a 7\% chance of existing in \Lam CDM. Using the
mass and cosmology quoted in Ref.\cite{Foley:2011jx} I find $\grmgrz=0.06$ in
good agreement with their result. With a mass cut of
$M>10^{15}\Msun$,\footnote{This cluster is also in W11 (see Section
\ref{sec:apploneclust}). The mass quoted in Ref.\cite{Foley:2011jx} is larger
than in W11 for reasons explained in Ref.\cite{Foley:2011jx}.} the unbiased
value for this cluster is $\nbgrmgrz=0.5$. Although this cluster certainly is
not a common cluster, its existence does not provide any tension with \Lam CDM.
\subsubsection{\nblessf}
Despite the fact that \nbgrmgrz~can be used as an unbiased measure of the
rareness of extreme galaxy clusters there are other possible unbiased measures
that will be more justified in certain circumstances. The calculation of the
expected number of clusters in a region of the $(M,z)$ plane depends on the
mass-function, $f$, and the volume element. If it is assumed that the expansion
history
of the universe is exactly \Lam CDM but that the primordial perturbations
are different then $f$ will change, but the volume element will be unchanged.
For this situation, it is useful to define the contours, $M(z)$, of constant
rareness as
contours of constant $f$. The rareness, \nblessf, is then defined to be the
probability that at least
one cluster lies above the contour of constant $f$. 

This definition would remove the possibly unwanted effect of a cluster being
called
`rare' simply because it was found in a region of the $(M,z)$ plane that had a
small volume. A region of the $(M,z)$ plane with small $f$ will be seen as rarer
by \nblessf~than by \nbgrmgrz~because regions with larger $f$ but smaller volume
will not be included in the expectation value used to calculate \nblessf. If,
however, the volume element \emph{was} different to \Lam CDM predictions
\nblessf~would be less sensitive to this than \nbgrmgrz.
\subsubsection{$R_{>z}$}
Alternatively, it could be assumed that the primordial perturbations are well
described by
a Gaussian power law but there are deviations from the \Lam CDM volume element
or growth function. In this case, the deviations
from \Lam CDM will depend only on redshift and not on mass at all. For this
possibility, it is useful to introduce a measure of rareness that quantifies
rareness only by how large $z$ is. Clusters found at large
redshift and small mass would be classified as rarer by this measure than
they would by other measures.

\subsubsection{Measures applied to existing extreme clusters}
\label{sec:apploneclust}
\begin{table}
\begin{tabular}{|c||c|c|c|c|}\hline
 \ph{0} Cluster  \ph{0} & \ph{00}\nblessf\ph{00} &\ph{00}\nbgrmgrz\ph{00}
&\ph{00} $<f$~ Mass at $z=0$\ph{00} & \ph{00}$>M>z$~ Mass at $z=0$\ph{00}\\
[1ex]\hline \hline
 \ph{0}J2235.3+2557 (H10)\ph{0} & $0.58$ & $0.49$&
$7.7\times10^{15}\Msun$&$3.3\times10^{15}\Msun$\\
[0.5ex]\hline
 J0546-5345 (H10)& $0.76$ & $0.61$& $6.2\times10^{15}\Msun$
&$2.8\times10^{15}\Msun$\\
[0.5ex]\hline
 J0910+5422 (H10) & $0.86$ & $0.79$& $4.5\times10^{15}\Msun$
&$1.8\times10^{15}\Msun$\\
[0.5ex]\hline
 J2215.9-1738 (H10) & $0.85$ & $0.81$&
$5.2\times10^{15}\Msun$&$1.8\times10^{15}\Msun$\\
[0.5ex]\hline\hline
J0102-4915 (W11) & $0.63$ & $0.61$&
$7.1\times10^{15}\Msun$&$3.8\times10^{15}\Msun$\\
[0.5ex]\hline
 J0615-5746 (W11) & $0.63$ & $0.70$& $7.1\times10^{15}\Msun$
&$3.5\times10^{15}\Msun$\\
[0.5ex]\hline
 J0658-5556 (W11) & $0.84$ & $0.63$& $5.2\times10^{15}\Msun$
&$3.6\times10^{15}\Msun$\\
[0.5ex]\hline
 J2106-5844 (W11) & $0.73$ & $0.86$&
$6.7\times10^{15}\Msun$&$3.0\times10^{15}\Msun$\\
[0.5ex]\hline
 J2248-4431 (W11) & $0.84$ & $0.66$&
$5.3\times10^{15}\Msun$&$3.5\times10^{15}\Msun$\\
[0.5ex]\hline
J2344-4243 (W11) & $0.92$ & $0.88$&
$5.0\times10^{15}\Msun$&$2.7\times10^{15}\Msun$\\
[0.5ex]\hline
\end{tabular}
\caption{\small Two unbiased rareness measures applied to the most extreme
clusters in H10 and W11. As well as, each cluster's `equivalent mass at redshift
zero'
according to both rareness measures. This is the mass of a cluster at redshift
zero that would be judged to be equally as rare as the actual observed cluster.
The value of $R$ should be interpreted as the probability that at least one
cluster could exist that is at least as rare as the observed cluster.}   
\label{tab:singclust}
\end{table}
I have applied these rareness measures to the clusters in H10 and
Ref.\cite{Williamson:2011jz} (hereafter W11). For the H10 clusters I have used
the same redshift cut, mass cut and \fsky~as in Section \ref{sec:multclustbias}.
If lower masses were allowed, then every cluster would become less rare because
more space would be opened up in the $(M,z)$ plane to find clusters. Even with
this cut in mass, none of the clusters are exceptionally rare. For the W11
clusters there is no cut in redshift, but a cut of $M>10^{15}\Msun$ and
$\fsky=2500$ sq. degrees. This cut is used to approximate the selection function
of W11.
W11 saw 26 clusters. With this mass
cut, the expected number of clusters observed is $>100$. In a more accurate
analysis, taking the selection function properly into account, the W11
clusters would be rarer than what I quote here. However, the same calculation
with a
higher mass cut, designed to produce only 26 expected clusters, gives
qualitatively the same results; that is, none of the clusters are exceptionally
rare. I have presented results for $M>10^{15}\Msun$ because this is the lowest
mass of any of the clusters observed in W11.

Table \ref{tab:singclust} lists \nblessf~and
\nbgrmgrz~for the most extreme H10 and W11 clusters. Uncertainties in the masses
of the
clusters have been accounted for with the method described in Section
\ref{sec:multclustbias}.\footnote{For the W11 clusters I added the
experimental and sytematic errors in quadrature.} In this table I have also
presented the ``equivalent mass at redshift zero'' of each cluster. As seen in
the previous section, an unbiased measure of rareness is constructed from a set
of contours in the $(M,z)$ plane defined to be equally rare. The ``equivalent
mass at redshift zero'' is then the point at which the $M(z)$ contour for a
given cluster and measure crosses zero. That is, another
cluster observed at redshift zero, with mass $M(0)$ would be judged to be
equally
rare. For these values I have chosen not to marginalise over the mass
uncertainties, using only the central value of
mass quoted in H10 and W11. The actual equivalent mass at redshift zero depends
on the rareness measure. The more intuitively sensible values are those for the
\nbgrmgrz~measure. The \nblessf~measure,
while useful when looking for new physics, loses something here because it
neglects the volume element, which is small at $z=0$. While some of these
clusters would be very massive at $z=0$, they would not be beyond
precedent\footnote{Care should be taken when comparing the masses for H10 and
for W11. Because of the different cuts in mass and redshift, equal values of $f$
or $\grmgrz$, for clusters from different sets, would \emph{not} equate to equal
values of \nblessf~and \nbgrmgrz.}.

Although no single cluster in Table \ref{tab:singclust} is extremely rare, it
might be thought that the total probability of having this many clusters with
rareness $\lesssim0.7$ would be very small. However, caution must be taken when
estimating the rareness of a set of clusters by looking at the rareness of its
constituents, especially when mass uncertainties are large. For any cluster,
large mass uncertainties will push $R$ towards 0.5 because the uncertainties
will make the cluster more consistent with being both very rare ($R\simeq0$) and
very generic ($R\simeq1$). When considering a set of clusters, the mass
uncertainties need to be taken into account all at once. If the uncertainties
are large then the same effect will drive $R$ towards 0.5 for the set as well.
Thus, two individual $R$'s close to 0.5 should not necessarily be interpreted as
meaning the two clusters are rare as an ensemble. It is possible that they are
both just not well measured. Nevertheless, two clusters with very small mass
uncertainties and rareness measures of $\sim0.5$ would be seen to be much rarer
as an ensemble. The following section will answer this question decisively for
H10 and W11. Neither set is particularly rare.

\subsection{Multiple cluster measures}\label{sec:manyclust}
Studying the rareness of individual galaxy clusters is useful and may provide
information about new physics. However it may be the case that any deviations
from \Lam CDM are subtle. Thus, no individual galaxy cluster might be rare
enough on its own to rule out \Lam CDM to a high degree of significance.
Perhaps, if there are a number of clusters, which individually provide only
small tension with \Lam CDM, they might collectively provide much more tension.
To quantify this it is necessary to have a measure of the rareness of a set of
clusters.

Quantifying the rareness of a set of clusters risks potential biases similar to
the individual rareness biases. Consider the situation where only two clusters
are being examined, clusters $A$ and $B$. An intuitive rareness measure,
$\tilde{R}^T$, is constructed by asking ``what is the probability of at least
one cluster existing that is at least as rare as $A$ and at least one cluster
existing that is
at least as rare as $B$?''. This measure would be biased. If $A$ was
made less rare but $B$ was made more rare to compensate then $\tilde{R}^T$
can remain the same. Therefore, the true probability of a pair of clusters
existing that are as rare as $A$ and $B$ is necessarily greater than
$\tilde{R}^T$. With sets of clusters care must also be taken because the
rareness of one cluster is not independent of the rareness of other clusters in
the set. Nevertheless, these obstacles can be overcome. I present below three
possible methods.
\subsubsection{$\tilde{R}^H$}
$\tilde{R}^H$ was introduced in Section \ref{sec:multclustbias} where it was
biased because I used it with \grmgrz. It would also be biased if instead I used
\nbgrmgrz. This second bias results from the method of multiplying the
individual cluster rareness together. Distributions that are uniform between
zero and one, will not give a uniform distribution when multiplied together. The
benefit of this measure is that it is easy to calculate. Any result can be
compared to sample cluster populations to check expectations and thus remove the
bias.
\subsubsection{$R^M$}
Another simple, multiple cluster measure of rareness is suggested in
Ref.\cite{Mortonson:2010mj}. This measure asks what is the probability
that there are at least $i$ clusters at least as rare as the $i$th rarest
cluster in the set. This gives
\begin{equation}\label{eq:R^M}
R^M_i=1-e^{-\lambda_i}\sum_n^i\frac{\lambda_i^n}{n!},
\end{equation}
where $\lambda_i$ is the expected number of clusters above the contour of
constant rareness for the $i^\mathrm{th}$ rarest cluster. $R^M$ is unbiased
because it
depends on only the one parameter, $\lambda_i$.
\subsubsection{$R^C$}
It could occur that all the $R^M_i$ are small but not small enough to claim a
significant deviation from \Lam CDM. Alternatively, one of the $R^M_i$ could be
very small, but all the rest $\gtrsim0.1$. What is needed is a way to
quantify the combined significance of \emph{all} of the $R^M_i$.
$\tilde{R}^H$ will do this to a certain degree, but crudely. Most of the
effect on $\tilde{R}^H$ comes from the one or two most extreme clusters
whereas $R^M_i$ could be significant at large $i$.

The easiest suggestion is to multiply each of the $R^M_i$~together. This is fine
and would work (it is biased, but the bias is easily accounted for). However, the individual $R^M_i$ measures are not independent of
each other. If $R^M_1$ is known to be very small then this enhances the
probability that $R^M_2$ will also be small. If the two individual measures are
naively multiplied together then two sets of clusters that are not equally rare
will be classified so. This weakens the sensitivity of the measure.

The solution to this is to use a measure defined as the probability of the
intersection of the rareness measures.
\begin{equation}
 \tilde{R}_i^C=P(R^M_1<R^M_{1(\rm{obs})}\cap
R^M_2<R^M_{2(\rm{obs})}\cap\dots\cap
R^M_i<R^M_{i(\rm{obs})}),
\end{equation}
where $R^M_{i(\rm{obs})}$ is the observed value of $R^M_i$. By definition,
$R^C_1=R^M_1$. The $i=2$ calculation
proceeds as follows. First, define $N_i$ to be the number of clusters, in a
sample universe, above the
contour of constant rareness associated to the observed $i^\mathrm{th}$ rarest
cluster. Define
$N_{ij}$ to be the number of clusters between the $i^\mathrm{th}$ and
$j^\mathrm{th}$ contours. Then
\begin{eqnarray}\nonumber
 \tilde{R}_2^C&=&P(N_2\geq2\cap N_1\geq1)\\ \nonumber
&=&P(N_{12}\geq0)\times P(N_1\geq2)+P(N_{12}\geq1)\times P(N_1=1) \\ \nonumber
&=&P(N_1\geq1)+P(N_{12}=0)\times P(N_1=1)  \\
&=&R^C_{1}-\lambda_1 e^{-\lambda_2},
\end{eqnarray}
which is true because $N_{12}=N_2-N_1$ and $P(N\geq n)=1-P(N<n)$. Similar
calculations for $\tilde{R}^C_3$ and $\tilde{R}^C_4$ return
\begin{eqnarray}\label{eq:R^C}\nonumber
 R^C_1&=&1-e^{-\lambda_1}, \\ \nonumber
\tilde{R}^C_2&=&R^C_1-\lambda_1e^{-\lambda_2}, \\ \nonumber
\tilde{R}^C_3&=&\tilde{R}^C_2-\lambda_1e^{-\lambda_3}\left(\lambda_2-\frac{
\lambda_1}{2}\right), \\
\tilde{R}^C_4&=&\tilde{R}^C_3-e^{-\lambda_4}\left(\frac{\lambda_1^3}{6}
+\lambda_1\lambda_2\lambda_3-\frac{ \lambda_1\lambda_2^2}{2}
-\frac{\lambda_3\lambda_1^2}{2} \right).
\end{eqnarray}
These rareness measures are of course biased because they depend on more than
one parameter. The bias can be removed by comparing observations to sample
cluster populations. It can also be removed by a much more efficient
calculation. It is actually only the $\lambda_i$ that we need to
know the distributions for and not the full $(M,z)$ distribution of all
clusters in the sky. The distribution for $\lambda_1$ is easily constructed
because the distribution of $R^M_1=1-\exp(-\lambda_1)$ is uniform. It might be
hoped that all of the $\lambda_i$ could be constructed from the corresponding,
uniform $R^M_i$ distribution. If all that was needed was the distribution of one
$\lambda_i$ on its own this would be fine. However, what is actually needed is
the conditional distribution for $\lambda_i$ given
the values of all the other $\lambda_j$ with $j\neq i$. These distributions can
be sampled with the following method. First, allocate a value to $\lambda_1$
using a uniform distribution for $R^M_1$. This assigns the contour of
the rarest cluster. Next it is necessary to assign the contour of the second
rarest cluster. The probability of observing at least one cluster between the
rarest contour and the second rarest contour is
\begin{equation}
 R^M_{12}=1-e^{-\lambda_{12}},
\end{equation}
where $\lambda_{12}=\lambda_2-\lambda_1$. A value can be allocated to
$\lambda_{12}$ using a uniform distribution for $R^M_{12}$. This determines
$\lambda_2$ and assigns the contour of the second rarest cluster. The same
process will work for every $\lambda_i$ by first determining $\lambda_{i-1}$ and
then allocating a value to
$\lambda_{i(i-1)}=\lambda_i-\lambda_{i-1}$ using the uniform distribution,
$R^M_{i(i-1)}=1-\exp(-\lambda_{i(i-1)})$. This method is much more efficient
than simulating an entire cluster population
(with $>10^4$ bins) and then finding the $i$ rarest clusters. Using the
distributions for $\lambda_i$, the bias in $\tilde{R}^C$ is removed by
calculating the probability that $\tilde{R}^C$ will be less than the value
observed.
\begin{equation}
 R^C=P(\tilde{R}^C<\tilde{R}^C_{(\rm{obs})}).
\end{equation}
For $i\leq4$, I have tested this method for calculating $R^C_i$ by applying it
to sample cluster populations. It is unbiased and produces a uniform
distribution.
\subsubsection{Measures applied to observations}\label{sec:multclustnobias}

Although none of the individual clusters in H10 and W11 are
significantly rare I speculated in Section \ref{sec:apploneclust} that perhaps
the sets as a whole might be. It is now possible to check. In Table
\ref{tab:manyclust}, I have listed $R^M_{i>M>z}$ and $R^M_{i<f}$ for the W11
clusters. I have used the method for accounting for mass uncertainties described
in Section \ref{sec:multclustbias}. To clear up a
possible ambiguity, I sample once from each mass distribution in the entire W11
set, I
then calculate the individual rareness of every cluster, pick the $i$ rarest and
then calculate the combined rareness. I do not pre-assign, before making each
sample, which are the rarest clusters. I use the same mass cut, redshift cut and
$f_\mathrm{sky}$ as I did for the W11 entries of Table
\ref{tab:singclust}.
\begin{table}
\begin{tabular}{|c||c|c|c|c|c|c|c|c|c|c|}\hline
 \ph{000} $i$  \ph{000} &\ph{00} 1\ph{00}
&\ph{00}2\ph{00}&\ph{00}3\ph{00}&\ph{00}4\ph{00}&\ph{00}5\ph{00}&\ph{00}6\ph{00}
&\ph{00}7\ph{00}&\ph{00}8\ph{00}&\ph{00}9\ph{00}&\ph{0}10 \ph{0}\\[1ex]\hline
\hline
 \nblessfi & $0.32$ &
$0.30$&$0.33$&$0.43$&$0.57$&$0.72$&$0.83$&$0.90$&$0.94$&$0.97$\\
[0.5ex]\hline
 \nbgrmgrzi & $0.22$ &
$0.21$&$0.23$&$0.28$&$0.36$&$0.44$&$0.53$&$0.62$&$0.71$&$0.78$\\
[0.5ex]\hline
\end{tabular}
\caption{\small The rareness of the $i^\mathrm{th}$ rarest cluster in W11, using
two different methods to define the contours of
constant rareness in the $(M,z)$ plane. The value of $R$ should be interpreted
as the probability that the $i^\mathrm{th}$ rarest existing cluster is at least
as rare as the $i^\mathrm{th}$ rarest W11 cluster.}   
\label{tab:manyclust}
\end{table}

For the first few values of $i$, both contour definitions indicate that the
rarest W11 clusters are slightly rarer than the \Lam CDM
average, but not with any statistical significance - there is at least a 20\%
chance of a set of clusters as rare as W11 occurring in \Lam CDM. As $i$
increases, $R$ approaches 1 for both
contour definitions. This indicates that, for the mass cut I used and for
$i\gtrsim 6$, the $i^\mathrm{th}$ rarest clusters in W11 are not as rare as they
should be. This could be an indication of new
physics; however it is much, much more likely to be the result of neglecting
selection function effects. I have already noted that with the mass cut I use,
more than
100 clusters are expected to exist. There are only 26 clusters in W11. SPT do
not expect to see every cluster that exists above the mass cut I have imposed.
Therefore, it is probable that the expected but missing rare clusters do exist,
but are near this mass threshold, rather than absent due to reasons of
fundamental physics. The reason that selection
function effects are more pronounced for larger values of $i$ is that larger
mass
clusters are both more likely to be seen and more rare. Therefore, it
is more likely that the rarest cluster in any survey has been seen than, for
example, the 5th rarest cluster.

In Table \ref{tab:manyclust2}, I have listed $R^C_{i<f}$, with $i\leq4$, 
for both W11 and H10. The method to account for mass uncertainties is exactly
equivalent to that for
Table \ref{tab:manyclust} and the mass cut, redshift cut and $f_\mathrm{sky}$ is
the same for each set as it was for Table \ref{tab:singclust}. Table
\ref{tab:manyclust2} points to the same conclusion as Table \ref{tab:manyclust}.
The
total significance of the four rarest clusters shows that sets of clusters as
rare as H10 and W11 would occur in $\gtrsim 30\%$ of \Lam CDM universes. This
could mean that we live in a \Lam CDM universe, or simply that the mass errors
in both sets are too large to distinguish any new physics. 
\begin{table}
\begin{tabular}{|c||c|c|c|c|}\hline
 \ph{00} $i$  \ph{00} & \ph{00}1\ph{00}
&\ph{00}2\ph{00}&\ph{00}3\ph{00}&\ph{00}4\ph{00}\\
[1ex]\hline\hline
 W11 & $0.32$ & $0.30$&$0.30$ &$0.31$\\
[0.5ex]\hline
 H10 & $0.29$ & $0.31$&$0.33$ &$0.36$\\
[0.5ex]\hline
\end{tabular}
\caption{\small The combined rareness, $R^C_{i<f}$, of the $i$ rarest clusters
in H10 and
W11. The value of $R$ should be interpreted as the probability that a set of $i$
clusters could exist that is at least as rare as the $i$ rarest clusters in H10
or W11.}   
\label{tab:manyclust2}
\end{table}

\begin{figure}
\centering
\includegraphics[height=0.25\textheight]{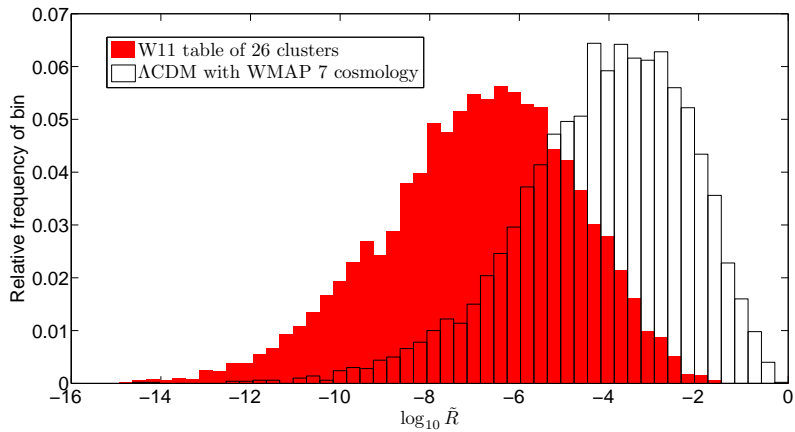}
\caption{\small Histogram of the rareness measure introduced in H10 applied to
both the W11 clusters and sample \Lam CDM populations. This is equivalent to
Figure \ref{fig:hoyleplot}, but for W11 instead of H10.} 
\label{fig:SPTplot}
\end{figure}

Finally, in Figure \ref{fig:SPTplot}, I have plotted histograms of
$\tilde{R}^H_{26>M>z}$~for
W11 and for sample cluster populations with the same mass
cut and $f_\mathrm{sky}$ as I have used for W11. The median value for
W11 is $\tilde{R}=1.7\times10^{-7}$. This value or less occurs for $\sim0.3$ of
the simulated populations. This is
consistent with what was seen using the other rareness measures in Tables
\ref{tab:manyclust} and \ref{tab:manyclust2}. The equivalent figure
for H10 is Figure \ref{fig:hoyleplot}. Figure \ref{fig:hoyleplot}
seems to indicate the H10 clusters are not rare enough for \Lam CDM, whereas
Table
\ref{tab:manyclust2} indicated that they were slightly too rare. Note however
that the $i$ values in Table \ref{tab:manyclust2} peak at
$i=4$, whereas Figure \ref{fig:hoyleplot} is for all 13 clusters in H10. As
already discussed, it is very unlikely that the H10 ensemble
really has seen the 13th rarest cluster in its observation window, whereas it is
quite possible that it \emph{has} seen the rarest four. An equivalent figure to
Figure \ref{fig:hoyleplot} including only the four rarest clusters in H10
and the sample cluster populations returns a result consistent with Table
\ref{tab:manyclust2}. This
effect probably was not seen in the comparable W11 analysis because of its more
complete survey\footnote{W11 has 26 of the $\sim 100$ clusters expected after my
choice of mass
cut - H10 has only 13 of the $\sim 300$ expected. Note that even in Table
\ref{tab:manyclust2} an increase in $R^C_i$ is noticeable for $i=4$.}.

To go further in the analysis would require modelling of the
selection function for each set of clusters. Given that both sets are slightly
rarer than a \Lam CDM average and the fact that accounting for the selection
function could only make the observed clusters appear rarer, this might be an
interesting exercise. Although, accounting for cosmological
uncertainties would weaken the strength of any obtainable result. To make an
estimate of the effects of selection functions I have attempted
re-doing the analyses presented here with a larger mass cut for W11. If I set
the mass cut to give an expected number of clusters of 26, the measured rareness
of W11 did become more significant, but $R^C$ did not fall below $0.05$.
However, if
selection functions are to be properly taken into account, any analysis might as
well go beyond just measuring the rareness of the most extreme clusters and
measure instead the total goodness
of fit of all the clusters observed. In the regions of sky considered by W11 and
H10, many more clusters have been observed than are actually analysed. This is
because both references made cuts ($z>1$ for H10 and significance of detection
($\sim$~mass) for W11) that isolate the most extreme clusters. The purpose of
measuring rareness is to find a
way to quantify the likelihood of these most extreme observed
clusters existing. It is less useful as a total goodness of fit measure for
100's of clusters.

\section{Rareness applied to primordial non-Gaussianity}\label{sec:nogauss}

As a case study of the use of rareness measures I will apply some of
the measures of the previous section to sample cluster populations
arising from non-Gaussian primordial density perturbations. No attempts have
been made to consider the effects of selection functions, cosmological
uncertainties or mass errors. The benefit of measuring the rareness of sample
populations is that this is not necessary. Therefore, although the results are
instructive of what would happen in a statistics limited experiment, they are
not representative
of any near-future experiments. A more accurate analysis of the constraints that
observations of a small
number of extreme galaxy clusters could provide on non-Gaussianity must be left
for
future work.

The method used to generate cluster populations from non-Gaussian primordial
density perturbations is identical to that used for Gaussian primordial density
perturbations except for the use of the mass function. I have chosen to
replace the Tinker et al. mass function used earlier with the prescription for a
non-Gaussian mass
function outlined in \cite{Paranjape:2011uk}. This prescription
provides a formula for the ratio between the full non-Gaussian mass function and
the Gaussian mass function. The prescription then dictates that the formula for
the ratio is to be multiplied by a mass function trusted in the Gaussian limit.
There are many prescriptions in the literature that each give slightly different
formulae for the ratio. To within the accuracy and range of masses, redshifts
and quantity of non-Gaussianity currently tested by N-body simulations, they
agree
with each other. I have chosen this particular prescription because it is
expected to be more stable (and therefore more accurate) at the larger masses,
redshifts and quantities of non-Gaussianity that have not yet been tested by
simulation. This
seems appropriate given that I am examining the most extreme objects in sample
cluster
populations. As in Ref.\cite{Paranjape:2011uk} and earlier sections of this
work, for the Gaussian limit I will use the Tinker et al. mass function. This
prescription parameterises the non-Gaussianity using the `local' template which
has the one parameter, \fnl \cite{Komatsu:2001rj}.

\begin{table}
\begin{tabular}{|c||c|c|c||c|c|c|}\hline
\multicolumn{1}{|c||}{}&\multicolumn{3}{|c||}{$\Cnblessf$}
&\multicolumn{3}{|c|}{$\nblessfi$}\\
[0.5ex]\hline
 \ph{0} $i$  \ph{0} & $\ph{00}\fnl=50\ph{00}$ &
$\ph{0}\fnl=100\ph{0}$ & $\ph{0}\fnl=500\ph{0}$ & $\ph{00}\fnl=50\ph{00}$ &
$\ph{0}\fnl=100\ph{0}$ & $\ph{0}\fnl=500\ph{0}$\\ [1ex]\hline \hline
 1 & $0.39$ & $0.30$& $0.022$		& $0.40$ &$0.30$ &$0.023$\\
[0.5ex]\hline
 2 & $0.37$ & $0.26$& $0.0050$		&$0.38$ &$0.24$ &$0.0024$\\
[0.5ex]\hline
 3 & $0.36$ & $0.24$& $0.0012$		&$0.34$ &$0.20$ &$3.0\times10^{-4}$\\
[0.5ex]\hline
 4 & $0.34$ & $0.21$& $3.1\times10^{-4}$&$0.32$ &$0.17$ &$4.5\times10^{-5}$\\
[0.5ex]\hline
\end{tabular}
\caption{\small The probability that the rarest clusters in a non-Gaussian
universe could exist in a Gaussian universe. $R^M_i$ quantifies the rareness of
the $i^\mathrm{th}$ rarest cluster. $R^C_i$ quantifies the combined rareness of
all of the $i$ rarest clusters.}   
\la{tab:fnl}
\end{table}
In Table \ref{tab:fnl}, I have calculated $R^M_{i<f}$ and $R^C_{i<f}$. To do
this I have used $f_\mathrm{sky}=1$ and made a mass
cut of $M>10^{14}\Msun$ and a redshift cut of $z>0.1$. I have generated sample
cluster populations using the quoted value of \fnl~and then asked
how rare each population would be in \Lam CDM. I have not included $\fnl=0$ in
the table because by construction it
returns $0.5$ for every $R$. Clearly, for large enough \fnl, observation of
even the rarest four galaxy clusters could rule out \Lam CDM. Disappointingly,
for smaller \fnl~this is not the case. Nonetheless, the two measures do behave
as expected. $R^C_i$ should quantify the total combined significance of the
the first $i$ $R^M_i$ measures. This seems to be
what is happening in the table. \nblessfi~drops as $i$ increases, indicating
that each individual statistic is further from \Lam CDM. \Cnblessf~drops
appropriately, although not as rapidly because it `remembers' the weaker
significance
of $R^M_j$ for all $j<i$.

The final result of this work is Figure \ref{fig:fnl100plot} which has
histograms of $\Hgrmgrz$ for $\fnl=100$ and \Lam CDM. This result matches what
is
seen in Table \ref{tab:fnl} for $\fnl=100$. This shows consistency amongst the
various rareness measures using different contours of constant rareness and
different means to combine
individual rareness. 

\begin{figure}
\centering
\includegraphics[height=0.25\textheight]{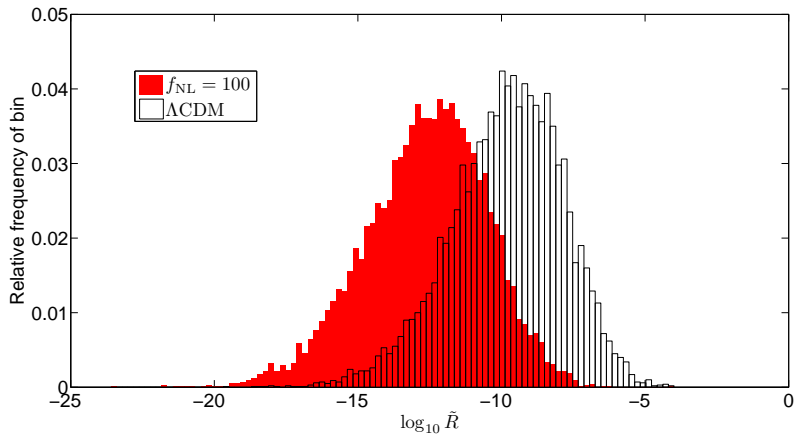}
\caption{\small Histograms of the rareness measure introduced in H10 applied to
a
population of Gaussian \Lam CDM universes and to a population of universes with
$\fnl=100$.} 
\la{fig:fnl100plot}
\end{figure}

\section{Summary and discussion}

In this work I have examined how to accurately quantify the rareness of extreme
galaxy clusters. I have shown that the most typical method, \grmgrz, is biased
and makes clusters appear less likely to exist than they actually are. The
easiest way to understand this is the fact that the probability of a cluster
existing with a given value of \grmgrz~is greater than \grmgrz~itself.
Therefore, the event of observing a cluster with a small value of \grmgrz~is not
necessarily expected to be an uncommon event. For example, in an \emph{actual}
\Lam CDM universe, there is a $\sim 40\%$ probability of a cluster existing that
would be claimed by \grmgrz~to rule out \Lam CDM at $90\%$ significance. I have
also shown this bias explicitly by calculating \grmgrz~for sample cluster
populations.

I have shown how to remove the bias in \grmgrz. I have also presented other
unbiased measures of the rareness of individual clusters, $R_{<f}$ and $R_{>z}$.
This set is not exhaustive of all possible measures. In fact, any prescription
that defines a unique set of $M(z)$ contours in the $(M,z)$ plane will define a
corresponding measure of rareness. For example, asking what the probability is
that a given cluster is the most massive cluster in a survey window would be
another possible rareness measure (see \cite{Cayon:2010mq}). I have argued that
$R_{<f}$ will be the measure that is most susceptible to deviations from \Lam
CDM arising through changes in the primordial sepctrum of density perturbations
and that $R_{>z}$ will be most susceptible to deviations in the expansion
history and growth function of the primordial perturbations.

I have presented methods to combine the individual rarenesses of a set of
clusters into a combined measure of the `rareness' of the set. $\tilde{R}_i^H$,
first used in Hoyle et al.\cite{Hoyle:2010ce}, simply multiplies together the
individual rarenesses of $i$ clusters. This method of combining rarenesses
introduces its own bias; however the bias can be straightforwardly accounted for
by comparing to sample populations. The most signficant flaw in $\tilde{R}^H$ is
that it is more sensitive to the rareness of the most extreme cluster, than it
is to any other cluster in a set. $R_i^M$, first used in Mortonson et
al.\cite{Mortonson:2010mj}, instead quantifies the probability that the
$i^\mathrm{th}$ rarest cluster in the universe could be as rare as the
$i^\mathrm{th}$ rarest in a given set. $R^M_i$ can then be calculated at each
value of $i$ to look for possible deviations from \Lam CDM. However, it could
occur that $R^M_i$ is small for all $i$, but never small enough to claim
significant tension, or $R^M_i$ could be significantly small for some values of
$i$, but not for others. $R^C_i$, first presented here, quantifies the combined
rareness of all of the $i$ rarest clusters. I have presented formulae for a
biased value of $R^C_i$ (i.e. $\tilde{R}^C_i$) for $i\leq4$ and a
computationally efficient method to account for this bias (without the need to
generate sample cluster populations). $R^C_i$ includes all the independent
information regarding rareness that is contained in the masses and redshifts of
the $i$ rarest clusters.

I have applied these individual and combined rareness measures to the sets of
clusters found in Ref.\cite{Hoyle:2010ce} (H10) and Ref.\cite{Williamson:2011jz}
(W11). The result is that each of the clusters has a greater than $50\%$
probability of existing in a \Lam CDM universe. There is also a greater than
$20\%$ probability that a set of clusters as rare as either H10 or W11 could
exist in \Lam CDM. These constraints are conservative and could be improved by
accounting for the selection function of each sample and reducing the mass
uncertainties. For the most extreme clusters in each sample I have also listed
the ``equivalent mass at redshift zero'', which is the mass of a cluster at
$z=0$ that would be judged to be equally rare. While some of the clusters would
be quite large at $z=0$ they would not be without precedent.

As a working example I applied the rareness measures to sample cluster
populations generated using a non-Gaussian primordial spectrum. For very large
non-Gaussianity, even the four rarest clusters could in principle rule out the
standard cosmological model with high significance. For more realistic levels of
non-Gaussianity, a small number of the most extreme clusters will not be able to
provide significant tension. Note however that my analysis has used the
non-Gaussianity parameter, \fnl. It is expected that the abundance of extreme
clusters (and voids) will be relatively more susceptible to higher order
non-Gaussianity parameters (e.g. \gnl, $\tau_\mathrm{NL}$) than \fnl~(see for
example Refs.\cite{Enqvist:2010bg,Chongchitnan:2010xz}).

In Ref.\cite{Mortonson:2010mj} convenient fitting formulae were provided for
$M(z)$ contours as a function of \fsky~and the exclusion limit with which one
desires to rule out \Lam CDM (using \grmgrz). Unfortunately, generalising these
contours to unbiased measures of rareness is not straightforward. The major
problem is that the region that it is necessary to integrate over in the $(M,z)$
plane to unbiasedly quantify rareness is experiment dependent. If an experiment
cannot see clusters below a certain mass threshold, a cluster observed by this
experiment should not be claimed to be `likely' because equivalently rare
clusters might exist below this threshold. A very conservative set of fitting
formulae could be developed by assuming that an experiment \emph{can} see
everything in the $(M,z)$ plane; however this would probably be so conservative
that it would effectively be impotent. Therefore, any accurate fitting formulae
would also need to depend on observational cuts made in the $(M,z)$ plane.

When calculating the rareness of observed galaxy clusters my aim was to show
that previously claimed sources of tension were a result of the use of the
biased measure \grmgrz. Sometimes, because \grmgrz~was so small, the previous
works in the literature made \emph{very} conservative assumptions elsewhere in
their analysis. I have also made these conservative assumptions. It is possible
that a less conservative analysis of the rareness of observed galaxy clusters
could still show significant tension with \Lam CDM. This is especially true of
the set of clusters in H10 (in constructing their table, H10 excluded clusters
without a spectroscopic redshift measurement, deliberately chose mass estimates
that had large errors, were conservative in their choice of \fsky~and neglected
any selection function effects). However, to avoid accidentally making false
claims of tension, any future analysis will need to either account for the bias
in \grmgrz, use one of the naturally unbiased measures presented here, or,
choose a new, well motivated, prescription for a set of equal rareness contours
in the $(M,z)$ plane.

\subsubsection*{Note:}

During the preparation of this manuscript, Ref.\cite{Jee:2011az} was posted to
the arXiv. In Ref.\cite{Jee:2011az}, the gravitational lensing masses of 27
high-redshift clusters are presented. Ref.\cite{Jee:2011az} claims to find
significant tension with \Lam CDM; however they do this using the biased
measures \grmgrz~and $\tilde{R}^H_{4>M>z}$. With a preliminary analysis, using $\fsky=100$ sq. degrees, I do not find that any
\emph{individual} cluster in Ref.\cite{Jee:2011az} provides tension with \Lam
CDM. Also, for the set of clusters in Ref.\cite{Jee:2011az}, I find
$R^C_{4<f}=0.19$ (and similar values for $i<4$). Tantalisingly, however, I find
the following sequence of values for $R^M_{i<f}$: $0.30$, $0.24$, $0.13$,
$0.072$,  $0.055$, $0.061$, $0.092$, $0.37$, $0.50$ and $0.58$. Unfortunately I
have not yet developed the machinery to evaluate $R^C_i$ for $i>4$ to test the
actual significance of this curious trend.

\section{Acknowledgements}
I thank Kari Enqvist, Chris Gordon, Ben Hoyle, Aseem Paranjape and Olli Taanila
for helpful comments relating to an earlier draft of this work. I also thank Ben
Hoyle for useful discussions during the early stages of this work and Syksy Rasanen for some late advice. The seed of
the idea for this work came from a post relating to Ref.\cite{Foley:2011jx} by
Fergus Simpson at the Cosmo Coffee website. I acknowledge the support of Academy
of Finland grant 131454.

\end{document}